\documentclass[aps,prl,twocolumn,showpacs,superscriptaddress,floats]{revtex4-1}
\usepackage{graphicx,bm,amsmath,amssymb,natbib,url,epsfig}

\begin{document}

\title{Anisotropic Charge Kondo Effect in a Triple Quantum Dot}
\author{Gwangsu Yoo}
\author{Jinhong Park}
\author{S.-S. B. Lee}
\author{H.-S. Sim}
\affiliation{Department of Physics, Korea Advanced Institute of
Science and Technology, Daejeon 305-701, Korea}

\date{\today}

\begin{abstract}
We predict that an {\em anisotropic} charge Kondo effect appears in a triple quantum dot, when the system has two-fold degenerate ground states of (1,1,0) and (0,0,1) charge configurations. 
Using bosonization and refermionization methods, we find that at low temperature,
the system has the two different phases of massive charge fluctuations between the two charge configurations and vanishing fluctuations, which are equivalent with the Kondo-screened and ferromagnetic phases of the anisotropic Kondo model, respectively. 
The phase transition is identifiable by electron conductance measurement, offering the possibility of experimentally exploring the anisotropic Kondo model.
Our charge Kondo effect has similar origin to that in a negative-$U$ Anderson impurity.
\end{abstract}
 



\maketitle

Kondo effects~\cite{Kondo,Hewson} and related quantum impurity problems are the central issues of low-dimensional many-body physics.
Many key features of Kondo effects have been explored in a controlled fashion, by utilizing quantum dots~\cite{Glazman,Goldhaber-Gordon,Cronenwett}.
A single quantum dot provides an ideal platform for studying the basic properties of Kondo effects, such as fractional shot noise~\cite{Sela,Yamauchi}, scattering phase shift~\cite{Ji,Takada}, and Kondo cloud~\cite{Affleck,Park}. Exotic Kondo effects~\cite{Jeong,Potok}
and pseudo-spin resolved transport~\cite{Amasha} were measured in a double dot. Triple quantum dots (TQDs) and larger systems are useful for artificially realizing magnetic effects~\cite{Rogge,Seo,Kuzmenko,Mitchell,Baruselli} including a geometrical frustration effect~\cite{Seo}.

The phase transition in the anisotropic Kondo effect, however, has not been experimentally explored in a controlled fashion.
It occurs between the Kondo-screened phase and the ferromagnetic coupling phase, and it is of Kosterlitz-Thouless type~\cite{Hewson}.
%
A single dot usually stays only in the Kondo phase. There have been the predictions of anisotropic Kondo effects in multiple dots~\cite{Kuzmenko,Mitchell,Baruselli,Garst,Romeike,Pletyukhov,Cornaglia}, but, they cover only a particular region of the phase diagram (e.g., only along the transition line~\cite{Kuzmenko,Mitchell,Baruselli}), or do not discuss experimental possibility.
It will be valuable to find a setup for observing the phase transition and the ferromagnetic phase.

\begin{figure}[tb]
\includegraphics[width=0.90\columnwidth]{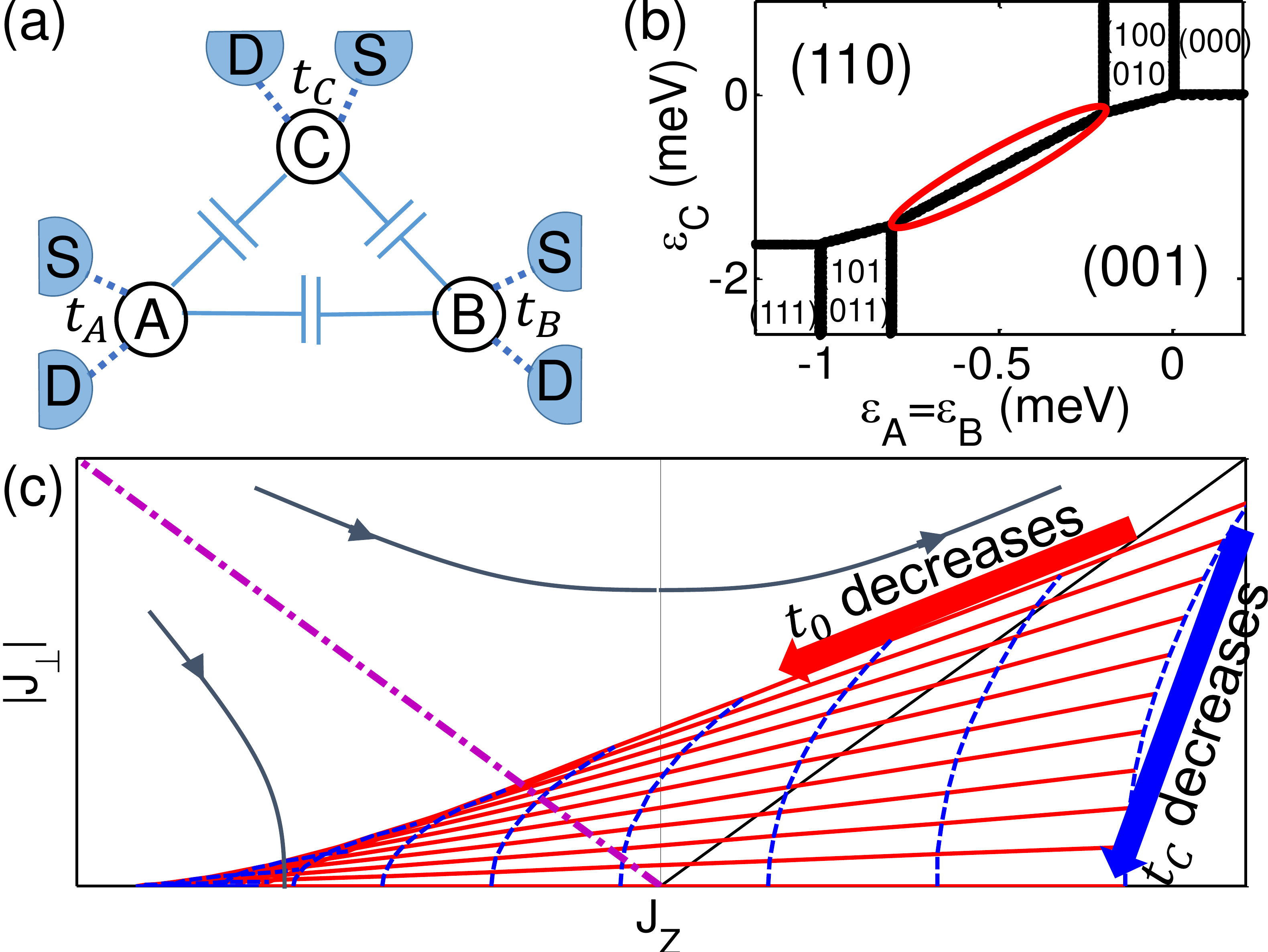}
\caption{
(Color Online) (a) Schematic view of TQD. It has no inter-dot electron tunneling, while each dot ($\lambda  = $ A,B,C) has tunneling strength $t_\lambda$ to its own source S and drain D. (b) TQD Stability diagram. The degeneracy line of (1,1,0) and (0,0,1) is marked. We choose intra-dot Coulomb repulsion 8 meV, inter-dot repulsion $U_\textrm{AB}=0.2$ meV, and $U_\textrm{0C} \equiv U_\textrm{AC} = U_\textrm{BC} = 0.8$ meV.
(c) Kondo phase diagram with renormalization group flow (thin arrows) and phase transition (dash-dotted magenta). 
The shaded region is the anisotropy domain achievable with tuning $t_0 \equiv t_\textrm{A} = t_\textrm{B}$ (along thick solid red arrow; one can tune only one of $t_\textrm{A}$ and $t_\textrm{B}$) and $t_\textrm{C}$ (dashed blue), starting from a location of $(J_z, J_\perp)$.
} \label{fig:model}
\end{figure}

In this work, we predict that an anisotropic charge Kondo effect appears in a TQD, when the system has the two-fold degenerate ground states of charge configuration $(n_\textrm{A},n_\textrm{B},n_\textrm{C}) = (1,1,0)$ and $(0,0,1)$; see Fig.~\ref{fig:model}. Here, $n_\lambda$ is electron occupation number in dot $\lambda=$ A,B,C. 
Using bosonization and refermionization~\cite{Zarand,vonDelft}, we find that the TQD has the two different phases of massive or vanishing fluctuations between the two 
ground-state charge configurations, which are equivalent with the two phases of the   anisotropic  Kondo effects.
Our Kondo effect of charge degrees of freedom is unusual, as the change (pseudospin flip) from one to the other ground state is accompanied by {\em three} electron-tunneling events, each involving different dots. This causes the anisotropy, which is experimentally controllable with changing the tunneling strengths.
Using numerical renormalization group (NRG) methods~\cite{Bulla}, we find that the two phases and the phase transition are experimentally accessible, and identifiable by electron conductance in a pseudospin-resolved fashion~\cite{Amasha}. This offers the possibility of experimentally exploring the anisotropic Kondo phase diagram.
Our charge Kondo effect is similar to that in a negative-$U$ Anderson impurity~\cite{Taraphder,KochSela}, and
has  different origin from those by orbitals~\cite{Amasha,Borda,Galpin}.
Note that the two-fold degeneracy was already observed~\cite{Rogge,Seo}.

{\it Model.---} The three dots of the TQD repulsively interact but have no inter-dot electron tunneling; see Fig.~\ref{fig:model}. Its Hamiltonian is $\mathcal{H} = \mathcal{H}_{\rm D} + \mathcal{H}_{\rm T} + \mathcal{H}_{\rm L}$. The dot part is
$\mathcal{H}_{\rm D} =\sum_{\lambda = \textrm{A,B,C}}\epsilon_\lambda n_\lambda + \sum_{\lambda \ne \lambda'} U_{\lambda \lambda'} n_\lambda n_{\lambda'}$.
Electron number operator $n_\lambda \equiv d_\lambda^\dagger d_\lambda$ is for the energy level $\epsilon_\lambda$ of dot $\lambda$ and $U_{\lambda \lambda'} > 0$ is inter-dot Coulomb energy; we consider one orbital per dot. Each dot $\lambda$ couples with its own two leads $\lambda \textrm{S}$ and $\lambda \textrm{D}$ via electron tunneling of strength $t_\lambda$, described by $\mathcal{H}_\textrm{T} = \sum_{\lambda,k} t_\lambda c^\dagger_{\lambda k}d_{\lambda} + \textrm{H.c.}$
$c^\dagger_{k \lambda}$ creates an electron of momentum $k$ and energy $\epsilon_k$ in $\lambda \textrm{S}$ and $\lambda \textrm{D}$; we consider the symmetric case of $c^\dagger_{\lambda k} = (c^\dagger_{\lambda \textrm{S} k} + c^\dagger_{\lambda \textrm{D} k})/\sqrt{2}$, $c^\dagger_{\lambda \textrm{S(D)} k}$ being an operator for $\lambda \textrm{S(D)}$.
The lead Hamiltonian is $\mathcal{H}_\textrm{L} = \sum_{\lambda,k}\epsilon_k c^\dagger_{\lambda k}c_{\lambda k}$; we omit the lead states decoupled from the TQD. For simplicity, we choose the intradot Coulomb interaction $\gg U_{\lambda \lambda'}$ (so double occupancy in dot level $\epsilon_\lambda$ is ignored), and the symmetric case of $\epsilon_\textrm{A} = \epsilon_\textrm{B}$, $U_\textrm{0C} \equiv U_\textrm{AC} = U_\textrm{BC}$, and $t_0 \equiv t_\textrm{A} = t_\textrm{B}$; relaxing the simplifications does not alter our results qualitatively, provided that the two-fold ground-state degeneracy is maintained. 
We ignore the spin of electrons and the ordinary spin Kondo effect in each dot, considering Zeeman energy (by a magnetic field) larger than the Kondo temperature of the spin Kondo effect~\cite{Costi2}. 


We impose the conditions for the two-fold degenerate ground states $(0,0,1)$ and $(1,1,0)$. 
One is $U_\textrm{AB}<U_\textrm{0C}$, making the energy of $(1,1,0)$ lower than that of the other double-occupancy states such as $(1,0,1)$. The others are  $-U_\textrm{0C} < \epsilon_\textrm{A/B} < -U_\textrm{AB}$, $-2U_\textrm{0C} < \epsilon_\textrm{C} < 0$, and $\epsilon_\textrm{A} + \epsilon_\textrm{B} -\epsilon_\textrm{C} = -U_\textrm{AB}$, achievable by gate voltages. 
For computational convenience, we consider the additional restrictions of $\epsilon_\textrm{A} = \epsilon_\textrm{B} = -(U_\textrm{AB} + U_\textrm{0C})/2$ and $\epsilon_\textrm{C} = -U_\textrm{0C}$, making the spectral function particle-hole symmetric. 
We plot the stability diagram in Fig.~\ref{fig:model}.

The two-fold degenerate ground states is considered as pseudospin states $|\Uparrow \rangle \equiv |n_\textrm{A} = 1, n_\textrm{B} = 1, n_\textrm{C} = 0 \rangle$ and $|\Downarrow \rangle \equiv |0, 0, 1 \rangle$. Then, a natural question is whether the TQD shows a charge Kondo effect~\cite{Taraphder}, massive charge fluctuations between the two states. We choose the pseudospin operators as $S_z = [n_\textrm{A} n_\textrm{B} (1-n_\textrm{C}) - (1-n_\textrm{A})(1-n_\textrm{B})n_\textrm{C}]/2$ and $S_+ = d_\textrm{A}^\dagger d_\textrm{B}^\dagger d_\textrm{C}$. Then, $\mathcal{H}$ is mapped onto
\begin{eqnarray}
&&\mathcal{H}_{\rm D} + \mathcal{H}_{\rm T} \Rightarrow \mathcal{H}_{0}  \label{eq:kondo-like hamiltonian}  \\ 
= &&\sum_{kk' \lambda} J_{z\lambda}S_z c_{\lambda k}^\dagger c_{\lambda k'}
+ \sum_{kk'k''} (J_{+}S_+c_{\textrm{C} k''}^\dagger c_{\textrm{B} k'} c_{\textrm{A} k} + \textrm{H.c.}), \nonumber \\
& & J_{z \textrm{A}} = J_{z \textrm{B}} = \frac{4t_0^2}{U_\textrm{0C}-U_\textrm{AB}} > 0,\quad \quad J_{z\textrm{C}} = -\frac{2t_\textrm{C}^2}{U_\textrm{0C}} < 0, \nonumber \\ 
& & J_+ = -8t_0^2t_\textrm{C} \left(\frac{1}{U_\textrm{0C}(U_\textrm{0C}-U_\textrm{AB})} + \frac{1}{(U_\textrm{0C}-U_\textrm{AB})^2}\right) \nonumber
\end{eqnarray}
by the Schrieffer-Wolff transformation~\cite{Hewson}. 
Contrary to the usual Kondo models of two species (spin up, down) of electrons, $\mathcal{H}_0$ has three species ($\lambda =$ A,B,C), and the $J_+$ term of the third order of $t_0^2 t_\textrm{C}$; the latter results in the anisotropy of our charge Kondo effect, as shown below. 



{\it Bosonization and refermionization.---} It is nontrivial whether $\mathcal{H}_0$ shows a Kondo effect.  To see this, we apply a bosonization method~\cite{Zarand,vonDelft,Supple}, where the field operator $c^\dagger_\lambda (0) = \sqrt{2\pi /L} \sum_k c^\dagger_{\lambda k}$ at $x=0$ (the position coupled to dot $\lambda$) in lead $\lambda$ is bosonized as $c^\dagger_{\lambda}(0) = e^{i \theta_\lambda} e^{i \phi_\lambda (0)}/\sqrt{a}$. Here, $e^{i \theta_\lambda}$ is the Klein factor of lead $\lambda$, $\phi_\lambda (x)$ are the bosonic field describing plasmon excitations in lead $\lambda$, $a$ is the short-distance cutoff, and each lead is treated as a one-dimensional wire of length $L$. $\mathcal{H}_0$ is bosonized,
\begin{eqnarray}
\mathcal{H}_0 &=& \frac{L}{2 \pi}\sum_{\lambda=\textrm{A,B,C}} J_{z\lambda}S_z(\partial_x\phi_\lambda(0) + \frac{2 \pi N_\lambda}{L})
+(\frac{L}{2 \pi a})^{3/2} \nonumber \\
& \times &
J_+S_+ e^{i\theta_\textrm{C}} e^{-i\theta_\textrm{B}} e^{-i\theta_\textrm{A}} e^{i(\phi_\textrm{C}(0)-\phi_\textrm{B}(0)-\phi_\textrm{A}(0))} + \textrm{H.c.}
 \nonumber \end{eqnarray}
In the first term, $\partial_x\phi_\lambda(0) + 2 \pi N_\lambda/L$ means electron density at $x=0$ in lead $\lambda$, where $N_\lambda$ is the total number of electrons in lead $\lambda$; we omit the normal ordering.

In the pseudospin flip $|\Uparrow \rangle \leftrightarrow |\Downarrow \rangle$, $N_\textrm{C} + (N_\textrm{A} + N_\textrm{B})/2$ and $N_\textrm{A} - N_\textrm{B}$ are conserved, as
$(N_\textrm{A}, N_\textrm{B}, N_\textrm{C})$ change by $(-1, -1, 1)$ or $(1,1,-1)$. Using this, we introduce pseudofermion numbers $N_\uparrow$, $N_\downarrow$, $N_x$. $N_{\uparrow (\downarrow)}$ counts the pseudospin-up (down) fermions that try to screen impurity pseudospin $\Downarrow$ ($\Uparrow$), while 
$N_x = r_1 [N_\textrm{C} + (N_\textrm{A} + N_\textrm{B})/2]$ counts the fermions irrelevant to the screening.
We choose $N_\uparrow + N_\downarrow = r_2 (N_\textrm{A} - N_\textrm{B})$, as the corresponding number is conserved in the spin-1/2 Kondo effect.
Here, $r_{1,2}$ are constants.
We choose $N_\uparrow - N_\downarrow = (2/3) (N_\textrm{A} + N_\textrm{B} - N_\textrm{C})$ from the analogy that $N_\textrm{A} + N_\textrm{B} - N_\textrm{C}$ changes by 3 in the pseudospin flip, while the corresponding number change is 2 in the spin-1/2 Kondo effect.
Hence,
\begin{eqnarray}
& & (N_\uparrow, N_\downarrow, N_x)^\intercal = M (N_\textrm{A}, N_\textrm{B}, N_\textrm{C})^\intercal, \label{eq:linear transformation} \\
& & M = \frac{2}{3}\left(\begin{array}{ccc} (1-l)/2 & (1+l)/2 & -\frac{1}{2} \\ -(1+l)/2 & -(1-l)/2 & \frac{1}{2} \\ \frac{1}{2} & \frac{1}{2} & 1 \end{array}\right)\textrm{  and  } l = \sqrt{\frac{3}{2}}, \nonumber
\end{eqnarray}
where $\intercal$ means matrix transpose.
We have chosen $r_{1,2}$ such that $M$ is proportional to a unitary matrix (which is $l M$). We define the Klein factors, $\exp (i \theta_{\uparrow,\downarrow,x})$,  and boson fields of the pseudofermions corresponding to $N_{\uparrow,\downarrow,x}$, 
\begin{eqnarray}
(\theta_\uparrow, \theta_\downarrow, \theta_x)^\intercal & = & l^2 M (\theta_\textrm{A}, \theta_\textrm{B}, \theta_\textrm{C})^\intercal, \nonumber \\
(\phi_\uparrow, \phi_\downarrow, \phi_x)^\intercal & = & l M (\phi_\textrm{A}, \phi_\textrm{B}, \phi_\textrm{C})^\intercal. \label{bosonTransformation}
\end{eqnarray}
%
The Klein factors are determined by commutators $[ \theta_\lambda, N_{\lambda'} ] = i \delta_{\lambda \lambda'}$, $\lambda \in \{\textrm{A,B,C} \}$ or $\{ \uparrow, \downarrow, x \}$, while the unitary matrix $l M$ is chosen for the bosons $\phi_{\uparrow,\downarrow,x}$, since the boson transformation should be unitary; the choice is justified by the following successful refermionization.

%

Using Eq.~\eqref{eq:linear transformation}, we write $\mathcal{H}_0$ as
\begin{eqnarray}
\mathcal{H}_0 &=& \frac{L}{2 \pi}\frac{1}{l}(J_{z\textrm{A}}-\frac{J_{z\textrm{C}}}{2})S_z\sum_{\sigma = \uparrow,\downarrow} w_\sigma (\partial_x\phi_\sigma(0) + l \frac{2 \pi N_\sigma}{L})\nonumber\\ 
& + & (\frac{L}{2 \pi a})^{3/2} e^{i\pi \frac{l-1}{2}} J_+ S_+e^{i\theta_\downarrow}e^{-i\theta_\uparrow} e^{il [\phi_\downarrow(0) - \phi_\uparrow(0)]} + \textrm{H.c.}, \nonumber \end{eqnarray}
where $w_\uparrow = 1$, $w_\downarrow = -1$, and we used $J_{z\textrm{A}} = J_{z\textrm{B}}$.
Here, we omit the term of $\frac{L}{2 \pi}\frac{1}{l}S_z(J_{z\textrm{A}} + J_{z\textrm{C}})(\partial_x \phi_x(0) + l 2 \pi N_x/L)$, which is marginal in Poorman's scaling (so it does not influence the Kondo effect of $\mathcal{H}_0$), since the pseudospin flip does not modify $N_x$ and $\phi_x$. 
%

We apply the Emery-Kivelson transformation~\cite{Zarand,vonDelft} of $U_\textrm{EK} =  e^{i \pi[ -(l/2) N_\uparrow + (2-l) N_\downarrow / 2]S_z} e^{i (l-1) S_z (\phi_\uparrow(0)-\phi_\downarrow(0))}$ to refermionize $\mathcal{H}_0 + \mathcal{H}_\textrm{L}$ by $\mathcal{H}_\textrm{K} = U_\textrm{EK} (\mathcal{H}_0 + \mathcal{H}_\textrm{L}) U^\dagger_\textrm{EK}$. This leads to the {\em anisotropic} Kondo Hamiltonian~\cite{Supple},
\begin{eqnarray}
\mathcal{H}_\textrm{K} & = & 2 J_z  S_z s_z + J_\perp ( S_+  s_-  + \textrm{H.c.}) + \sum_{k \sigma} \tilde{\epsilon}_{k \sigma} f^\dagger_{k \sigma} f_{k \sigma}, \nonumber \\
J_z &\equiv& \frac{1}{l}(J_{zA}-\frac{J_{zC}}{2}) - \frac{l-1}{\rho}, \, \,
J_\perp \equiv \sqrt{\frac{L}{2 \pi a}}J_+. \label{eq:s dot s hamiltonian coefficients} \end{eqnarray}
$f^\dagger_{k \sigma}$ creates a refermionized fermion with spin $\sigma$, momentum $k$, and energy $\tilde{\epsilon}_k$, $s_z$ and $s_\pm = s_x \pm i s_y$ are the spin operators of the fermions ($f_{k \sigma}^\dagger$) coupled to the TQD pseudospin, and $\rho$ is the density of states of leads $\lambda$.
The term $-(l-1)/\rho$ is contributed from the fermion spin effectively bound to the TQD pseudospin.

\begin{figure}[t]
\includegraphics[width=\columnwidth]{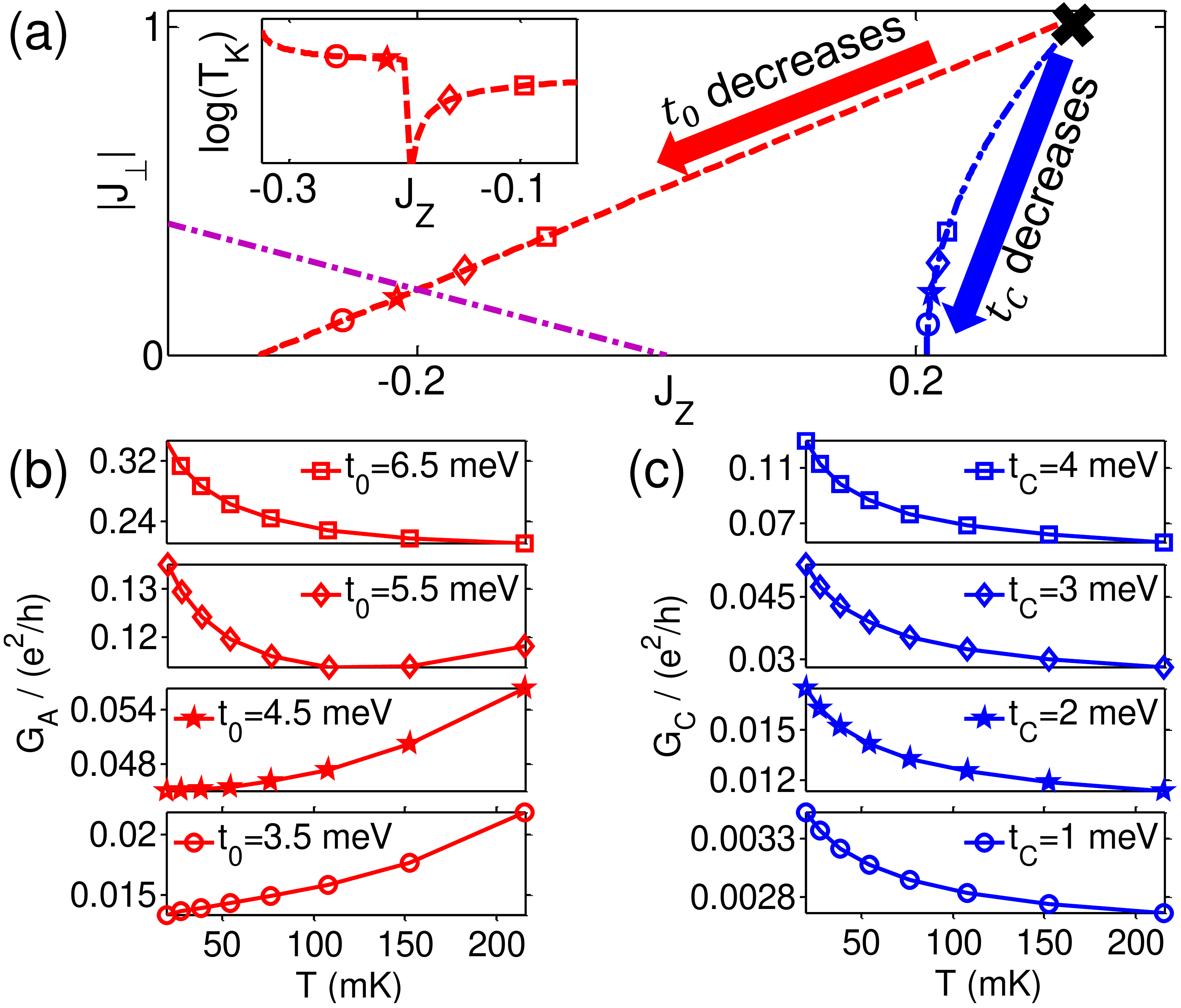}
\caption{
(Color online) NRG results of differential conductance $G_\textrm{A(C)}$ through dot A (C). 
(a) $(J_z, J_\perp)$'s, at which $G_\textrm{A(C)}$ is computed in (b) and (c), are marked by symbols. They are selected, starting from $t_0 = t_\textrm{C} = 11$ meV (marked by cross) and lowering $t_0$ (along red dashed trajectory) or $t_\textrm{C}$ (blue dashed-dot). The other parameters such as $U_{\lambda \lambda'}$ are the same with those in Fig.~\ref{fig:model}. In this choice, the trajectory crosses the phase transition (solid line) as $t_0$ decreases, while it stays within the Kondo phase as $t_C$ varies.
Inset: $T_K$ along the trajectory of lowering $t_0$. It shows a jump at the transition.
(b) The dependence of $G_\textrm{A}$ on $T$  at different $t_\textrm{0}$'s, with fixed $t_C=11$ meV along the red dashed trajectory of (a).
(c) $G_\textrm{C} (T)$ at different $t_\textrm{C}$'s, with fixed $t_0=11$ meV along the blue dash-dot trajectory of (a).
  In the ferromagnetic (Kondo-screened) phase, $G_\textrm{A,C}$ becomes smaller (larger) as $T$ decreases.
} 
\label{fig:conductance change} 
\end{figure}

In Eq.~\eqref{eq:s dot s hamiltonian coefficients}, $J_z$ can be negative, with the help of $- (l-1)/\rho < 0$. By tuning $J_z$ and $J_\perp$ (namely, $t_0$ and $t_C$; see Eq.~\eqref{eq:kondo-like hamiltonian}), it is possible to reach both the Kondo-screened and ferromagnetic phases, crossing the transition between them; see Fig.~\ref{fig:conductance change}(a). In the Kondo-screened phase, charge fluctuations (pseudo-spin flip) between the two ground states massively occur at low temperature, showing a charge Kondo effect, while they vanish in the ferromagnetic phase.

{\em Phase transition.---} To confirm the charge Kondo effect and the phase transition in our TQD, we apply NRG methods~\cite{Bulla} to the
initial Hamiltonian $\mathcal{H}$ (Anderson model), with the parameters close to recent experiments~\cite{Amasha}; we choose the bandwidth $D$ of the leads as $D = 1 \, \textrm{eV}$, and the NRG calculation is performed with spinless electrons. We compute differential electron conductance $G_\lambda$ through dot $\lambda$, applying bias $V \to 0$ to lead $\lambda$S; the other leads of $\lambda$D, $\lambda'$S, and $\lambda'$D ($\lambda' \ne \lambda$) are unbiased. 
Figure~\ref{fig:conductance change} shows $G_\textrm{A}$ and $G_\textrm{C}$. 
At $t_0 = t_\textrm{C} = 11$ meV [the corresponding $(J_z,J_\perp)$ is marked by cross in Fig.~\ref{fig:conductance change}(a)], the spectral function (not shown here) and $G_\lambda$ (see Ref.~\cite{Supple}) show that the TQD is in the Kondo phase and has Kondo temperature $T_K \gtrsim 150$ mK, which is achievable in current experiments. Here, $T_K = D [ (J_{z}+ \Delta_J)/(J_{z}-\Delta_J) ]^{-1/(4\rho \Delta_J)}$ and $\Delta_J = \sqrt{J_{z}^2-J_{\perp}^2}$;
$T_K \to D e^{- 1 / (2 \rho J_z)}$ in the $J_z = J_\perp$ limit.
For smaller $t_0$ and $t_\textrm{C}$,
the Kondo temperature determined by NRG methods decreases, following the above expression of $T_K$.

As $t_0$ decreases, with $t_\textrm{C} =  11$ meV fixed, the TQD moves to the ferromagnetic phase, crossing the phase transition. The transition is identified by the dependence of $G_\lambda$ on temperature $T$; see Fig.~\ref{fig:conductance change}. 
In the Kondo regime, $G_\lambda$ increases as $T$ decreases; $G_\lambda \propto 1-(\pi T/T_K)^2$ at $T \ll T_K$. 
In contrast, in the $T > T_K$ regime of the Kondo-screened phase (the red square and diamond in Fig.~\ref{fig:conductance change}) and the ferromagnetic phase (star and circle), we find the $T$ dependence~\cite{Supple},
\begin{eqnarray}
G_\lambda \propto \left(\Delta_J \frac{1+(T/T_{K})^{-4\rho \Delta_J}}{1-(T/T_{K})^{-4\rho \Delta_J}} + m_\lambda \right)^2 + n_\lambda
\label{eq:weak coupling regime conductance}
\end{eqnarray}
where $m_\lambda$ and $n_\lambda$ are $T$-independent; their expressions are found in Ref.~\cite{Supple}, and $G_\lambda = (1/\ln(T/T_{K}) + m_\lambda)^2 + n_\lambda$ when $J_z=J_\perp$~\cite{Glazman}. 
In the ferromagnetic phase, $G_\lambda$ increases as $T$ increases, showing the opposite behavior to the Kondo regime.
The ferromagnetic phase is also characterized by $T_K$, although $T_K$ has a different meaning from the Kondo phase, and has the spectral function $\propto [\ln (\omega / T_K)]^{-2}$ increasing with energy $\omega$~\cite{Kuzmenko,Mitchell,Baruselli,Koller};
the jump of $T_K$ at the transition is shown in the inset of Fig.~\ref{fig:conductance change}(a).
Around the transition, $G_\lambda$ shows the crossing behavior between the two opposite $T$ dependence.

On the other hand, as $t_C$ decreases, with $t_0 =  11$ meV fixed, the TQD stays in the Kondo screened phase, so that $G_\lambda$ increases as $T$ decreases. The $T$ dependence of $G_\lambda$ and $T_K$ at different $t_0$'s and $t_\textrm{C}$'s will be useful for experimentally studying the charge Kondo effect and the Kondo phase transition in the TQD. Note that as we choose the parameters of $\mathcal{H}$ close to experiments~\cite{Amasha},  $G_\lambda$ is affected by the tails of Coulomb blockade resonances (Hubbard satellites) of $\mathcal{H}$ in Fig.~\ref{fig:conductance change}; in Fig.~\ref{fig:conductance change}(b), $G_\textrm{A}$ shows nonzero value at $T \to 0$ in the ferromagnetic phase due to the tails, and we plot $G_\textrm{A}$, instead of $G_\textrm{C}$, as the former is less affected by the tails.

We note that a proper range of $t_0$ and $t_\textrm{C}$ for observing the phase transition is accessible in experiments, by using quantum point contacts formed in a two-dimensional electron system of a semiconductor heterostructure,  by selecting electron leads with a proper value of the density of states, or by selecting quantum dots with proper strengths of Coulomb interactions.


{\em Isotropic regime.---} In the Kondo phase, the TQD approaches, at low enough $T$, to the strong coupling fixed point of the isotropic Kondo effect. Applying Eq.~\eqref{bosonTransformation} to the standard fixed-point Hamiltonian~\cite{Nozieres,Supple}, we derive the effective fixed-point Hamiltonian of the charge Kondo effect of three electron species $\lambda=$A,B,C as
\begin{eqnarray}
\mathcal{H}_\textrm{fp} &=& \mathcal{H}_\textrm{L} - \frac{1}{2\pi\rho T_K} \sum_{\lambda=A,B,C;k k'}\alpha_\lambda  (\epsilon_k + \epsilon_{k'}) c^\dagger_{\lambda k} c_{\lambda k'} \nonumber\\
 & + & \frac{1}{\pi\rho^2 T_K}\sum_{\lambda'\neq\lambda;k_1 k_2 k_3 k_4}\beta_{\lambda\lambda'}c_{\lambda k_1}^\dagger c_{\lambda k_2}c_{\lambda' k_3}^\dagger c_{\lambda' k_4}, \label{eq:fixed point hamiltonian in terms of a, b, and c}
\end{eqnarray}
where $\alpha_\textrm{A}= \alpha_\textrm{B}= \alpha_\textrm{C}= 2/3$, 
 $\beta_\textrm{AB}=- 2/3$, and $\beta_\textrm{AC}=\beta_\textrm{BC}=2/3$.
The $\alpha_\lambda$ terms describe elastic scattering, while $\beta_{\lambda \lambda'}$ the inelastic scattering destroying the Kondo singlet.
The sign of $\beta_{\lambda \lambda'}$ carries the information whether the inter-dot interaction is repulsive or attractive. $\beta_\textrm{AC(BC)} > 0$ means the repulsive interaction between dot A (B) and C, while $\beta_\textrm{AB}<0$ interestingly implies that the interactions between A and B are effectively attractive. This stands in contrast to the usual Kondo fixed-point Hamiltonian of positive $\beta$'s. 

The effective attractive interaction between A and B is reminiscent of that of a negative-$U$ Anderson impurity~\cite{AndersonNU}. A negative-$U$ impurity is a site at which two electrons interact attractively with negative charging energy $U<0$,
and prefers zero or double electron occupancy, rather than single occupancy.
It results in the charge Kondo effect~\cite{Taraphder,KochSela} of electron-pair fluctuations between the zero and double occupancy, and electron-pair tunneling in molecules~\cite{Alexandrov,Cornaglia,Andergassen,Koch}.
In our TQD, the degenerate ground states prefer double or zero occupancy in dots A and B, as $(n_\textrm{A},n_\textrm{B}) = (1,1)$ or $(0,0)$, rather than single occupancy.
So, although the bare capacitive interactions of the TQD are repulsive,
the effective attraction of $\beta_\textrm{AB}<0$ arises between
electron species A and B in the TQD, with the help of dot C. 
This indicates that our charge Kondo effect of massive charge-pair fluctuations has similar origin to the charge Kondo effect in a negative-$U$ impurity.
Note that a negative-$U$ impurity shows only the isotropic Kondo-screened phase~\cite{Taraphder,KochSela}, contrary to our case.

Observables depend on $\alpha_\lambda$ and $\beta_{\lambda \lambda'}$ in the Kondo phase of the TQD, in a different way from usual Kondo effects. Below, we apply bias voltages $V_\lambda$ to leads $\lambda$S, satisfying 
$k_B T \ll V_\textrm{A}, V_\textrm{B}, V_\textrm{C} \ll k_B T_K$. From Eq.~\eqref{eq:fixed point hamiltonian in terms of a, b, and c} and following Refs.~\cite{Nozieres,Mora}, we obtain the phase shift of electrons of energy $\epsilon$ and species $\lambda$, scattered by the Kondo resonance,
\begin{eqnarray}
\delta_{\lambda} &=& \frac{\pi}{2} + \alpha_\lambda \frac{\epsilon}{T_K} - \sum_{\lambda'\neq\lambda} \beta_{\lambda\lambda'} \frac{e V_{\lambda'}}{T_K}. \label{eq:kondo phase shift}
\end{eqnarray}
From $d \delta_\lambda / d V_{\lambda'}$, one can experimentally~\cite{Ji,Takada} measure $\beta_{\lambda \lambda'}$ and confirm the attractive interaction of $\beta_{AB} < 0$.


We also derive, using Keldysh formalism~\cite{Sela}, electron current through dot $\lambda$ as [up to $O(V^3)$]
\begin{equation}
I_\lambda = \frac{e^2}{h} [ V_\lambda - (\frac{\alpha_\lambda^2}{12} + \sum_{\lambda' \ne \lambda} \frac{\beta_{\lambda \lambda'}^2}{6}) \frac{e^2 V_\lambda^3}{T_K^2} - \sum_{\lambda' \ne \lambda} \frac{\beta_{\lambda \lambda'}^2}{4} \frac{e^2 V_\lambda V_{\lambda'}^2}{T_K^2}]. \label{transconductance}
\end{equation} 
For $V_\lambda = 0$, $I_\lambda$ vanishes even if $V_{\lambda' \ne \lambda} \ne 0$. Using conductance $G_{\lambda} = dI_\lambda / dV_\lambda$ and transconductance $G_{\lambda\lambda'} = dI_\lambda / dV_{\lambda'}$, one can experimentally obtain $\alpha^2$ and $\beta^2$, the information of the scattering in the Kondo effects. Hence, our TQD is useful for studying the Kondo effect, by measuring the single-particle observable of $I_\lambda$ in a pseudo-spin resolved fashion~\cite{Amasha}.
Note that similar information can be obtained from shot noise~\cite{Sela,Yamauchi}, a two-particle observable.

{\it Summary.---}
We have shown that an anisotropic charge Kondo effect with tunable anisotropy appears in the TQD of the two-fold ground-state degeneracy of (1,1,0) and (0,0,1) charge configurations. Interestingly, this pseudospin-1/2 Kondo effect is accompanied by three different electron species (dot A, B, C) contrary to the usual Kondo effect of two species (spin up, down).
The TQD is useful for studying the Kondo phase transition, the ferromagnetic phase, the effective attractive interactions similar to those in a negative-$U$ Anderson impurity, 
and the inelastic Kondo scattering in a pseudospin-resolved fashion~\cite{Amasha}.
Note that the TQD is the minimal quantum dot system possessing a charge Kondo effect with effective attractive interactions. Larger systems such as a quadruple dot can also show a similar effect~\cite{Oreg}.  





We thank Yunchul Chung, David Goldhaber-Gordon, and especially Yuval Oreg for discussion. This work was supported by Korea NRF (Grant No. 2011-0022955, Grant No. 2013R1A2A2A01007327).

\end{document}


\title{Supplementary Material for ``Anisotropic Charge Kondo Effect in a Triple Quantum Dot''}
\author{Gwangsu Yoo}
\author{Jinhong Park}
\author{S.-S. B. Lee}
\author{H.-S. Sim}
\affiliation{Department of Physics, Korea Advanced Institute of
Science and Technology, Daejeon 305-701, Korea}

\date{\today}
\maketitle

In this Supplementary Material, we provide the expression of $\mathcal{H}_\textrm{L}$ in the bosonization, the details of the refermionization in Eq. (4), the dependence of $G_\textrm{A(C)}$ on $t_\textrm{0(C)}$, and the derivation of Eqs. (5) and (6) of the main text.


We first provide the expression of $\mathcal{H}_\textrm{L} = \sum_{\lambda k}\epsilon_k c^\dagger_{\lambda k}c_{\lambda k}$ in the bosonization, which is skipped in the main text. After the unitary transformation of Eq. (3), we have
\begin{eqnarray}
\mathcal{H}_\textrm{L} &=& \hbar v_F \sum_{\lambda = A,B,C}\int_{-L/2}^{L/2}\frac{dx}{2\pi}\frac{1}{2}(\partial_x\phi_\lambda(x))^2
 = \hbar v_F \sum_{\sigma = \uparrow,\downarrow,X}\int_{-L/2}^{L/2}\frac{dx}{2\pi}\frac{1}{2}(\partial_x\phi_\sigma(x))^2. \nonumber
\end{eqnarray}

Next, we discuss the refermionization details in the derivation of Eq. 
(4).
We apply the Emery-Kivelson transformation $U_\textrm{EK}$
to the total Hamiltonian of $\mathcal{H}_\textrm{L}+\mathcal{H}_{0}$ as  $U_\textrm{EK}(\mathcal{H}_\textrm{L}+\mathcal{H}_{0})U^\dagger_\textrm{EK}$. This results in
\begin{eqnarray}
U_\textrm{EK}(\mathcal{H}_\textrm{L} + \mathcal{H}_{0,z})U^\dagger_\textrm{EK}
 &=& \mathcal{H}_\textrm{L} + \mathcal{H}_{0,z} - \frac{(l-1)L}{2 \pi \rho} S_z \sum_{\sigma=\uparrow,\downarrow}w_\sigma\partial_x\phi_\sigma(0), \nonumber \\
U_\textrm{EK} \mathcal{H}_{0,\perp}U^\dagger_\textrm{EK}
 &=& (\frac{L}{2\pi a})^{3/2} J_+ S_+  F_\downarrow^\dagger F_\uparrow e^{i[\phi_\downarrow(0)-\phi_\uparrow(0)]}\nonumber
\end{eqnarray}
where $\mathcal{H}_0 = \mathcal{H}_{0, z} + \mathcal{H}_{0,\perp}$,
\begin{eqnarray}
\mathcal{H}_{0,z} &=& \frac{L}{2 \pi}\frac{1}{l}(J_{z\textrm{A}}-\frac{J_{z\textrm{C}}}{2})S_z\sum_{\sigma = \uparrow,\downarrow} w_\sigma (\partial_x\phi_\sigma(0) + l \frac{2 \pi N_\sigma}{L})\nonumber\\ 
\mathcal{H}_{0,\perp} & = & (\frac{L}{2 \pi a})^{3/2} e^{i\pi \frac{l-1}{2}} J_+ S_+e^{i\theta_\downarrow}e^{-i\theta_\uparrow} e^{il [\phi_\downarrow(0) - \phi_\uparrow(0)]} + \textrm{H.c.}. \nonumber \end{eqnarray}
The Klein factors in the EK transformed Hamiltonian are defined as
\begin{eqnarray}
F_\downarrow^\dagger &\equiv& e^{i(\theta_\downarrow -\frac{l-1}{2}\pi N_\uparrow)}, \nonumber \\
F_\uparrow^\dagger &\equiv& e^{i(\theta_\uparrow + \frac{l-1}{2}\pi N_\downarrow)}, \nonumber
\end{eqnarray}
satisfying $[N_\sigma, F_\sigma] = -F_\sigma$.
Phase operators are applied to $d_\lambda$ as $d_\lambda \to e^{-i \pi S_z/2} d_\lambda e^{i \pi (S_z + N_\uparrow - N_\downarrow)/2}$, to achieve the commutation relation of  $\{d_\lambda,F_\sigma\}=0 $. 
Here, 
$N_\sigma$ is the total number of pseudofermion in lead $\sigma$, and we have used the relation of $\rho=L/(2\pi \hbar v_F)$ in the computation of $U_\textrm{EK}(\mathcal{H}_\textrm{L} + \mathcal{H}_{0,z})U^\dagger_\textrm{EK}$.
Then, we are ready to refermionize the transformed Hamiltonian of $U_\textrm{EK}(\mathcal{H}_\textrm{L}+\mathcal{H}_{0})U^\dagger_\textrm{EK}$. We introduce the pseudofermion operator $f_\sigma (x=0)$ corresponding to the pseudofermion number $N_\sigma$ as
\begin{equation}
\makeatletter 
\renewcommand{\theequation}{S\@arabic\c@equation}
\makeatother
f_\sigma(0) \equiv \frac{F_\sigma}{\sqrt{a}}  e^{-i\phi_\sigma(0)}, \quad
f_\sigma^\dagger(0) f_\sigma(0) = \partial_x \phi_\sigma(0) + l\Delta_L N_\sigma. \label{F-operator}
\end{equation}
Combining this with $U_\textrm{EK}(\mathcal{H}_\textrm{L}+\mathcal{H}_{0})U^\dagger_\textrm{EK}$ and $F^\dagger_{\uparrow, \downarrow}$, we obtain
Eq. (4).

\begin{figure}[t]
\makeatletter 
\renewcommand{\thefigure}{S\@arabic\c@figure}
\makeatother
\includegraphics[width=0.5\columnwidth]{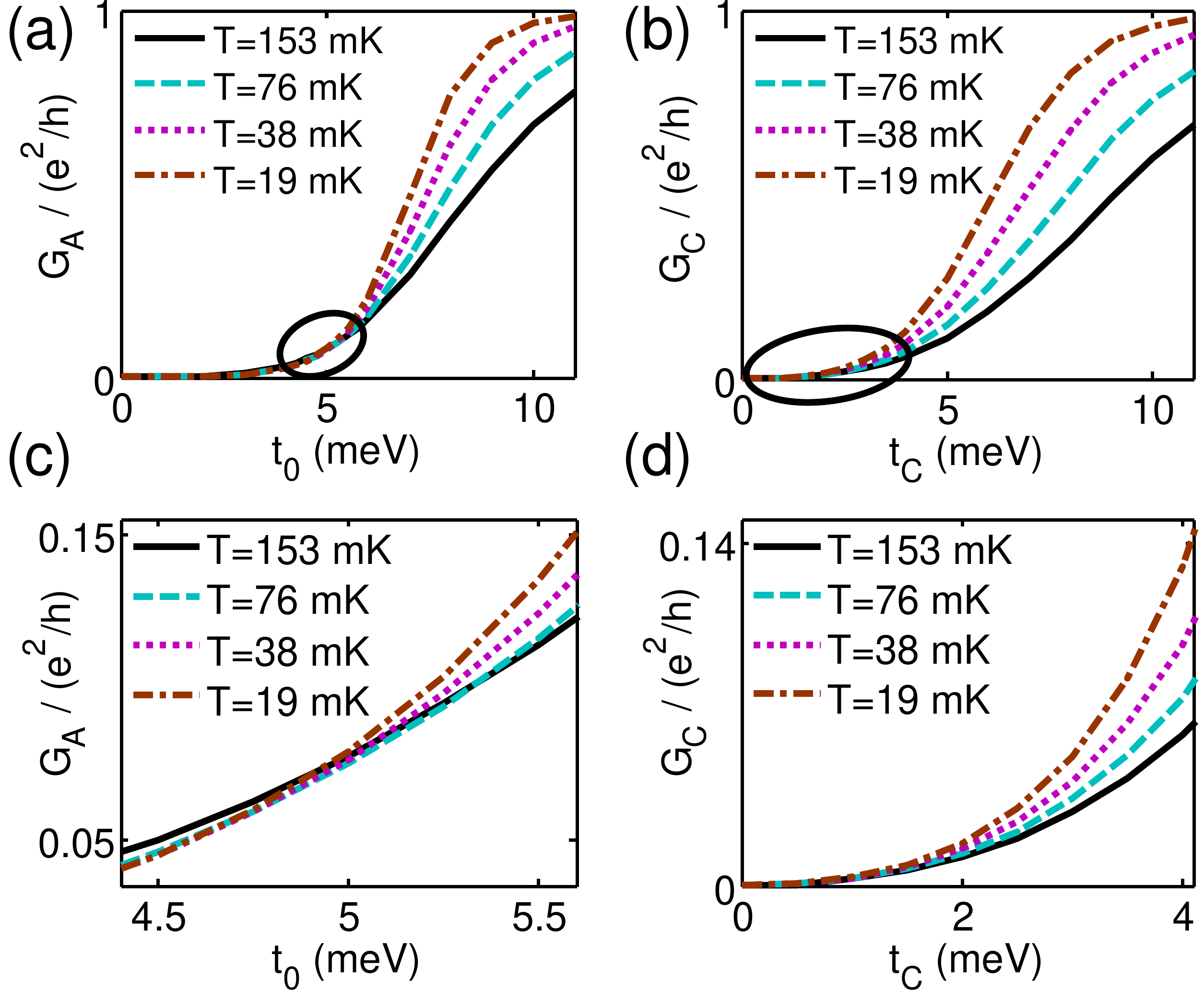}
\caption{
(a) Dependence of $G_\textrm{A}$ on $t_\textrm{0}$ at different temperatures $T$, with fixed $t_C=11$ meV, along the red dashed trajectory of Fig.~2(a).
(b) Dependence of $G_\textrm{C}$ on $t_\textrm{C}$ at different temperatures $T$, with fixed $t_0=11$ meV, along the blue dash-dot trajectory of Fig.~2(a).
The marked regions of (a) and (b) are magnified in (c) and (d), respectively.}
\label{fig:fig3} 
\end{figure}

Next, in Fig.~\ref{fig:fig3}, we plot the
the dependence of $G_\textrm{A(C)}$ on $t_\textrm{0(C)}$ at different temperatures. 
The differential conductance approaches to the unitary limit of $e^2/h$ at low temperature in the Kondo phase; we ignore the spin degree of freedom, considering a finite Zeeman energy. And, around $t_0 = t_C = 11$ meV, the Kondo temperature is obtained as about 150 mK, which is experimentally achievable. In Fig.~\ref{fig:fig3}(c), $G_\textrm{A}$ shows the crossing behavior around the transition between the antiferromagnetic and ferromagnetic phases, which is supplementary to Fig.~2(b).


Next, we derive Eq. (5). Conductance $G_\lambda$ is obtained, using t-matrix $\mathcal{T}_\lambda$, as~\cite{Glazman}
\begin{equation}
G_\lambda=G_0\int d\omega (-\frac{df}{d\omega})\frac{1}{2} [-\pi\rho \textrm{Im} \mathcal{T}_\lambda(\omega)], \nonumber
\end{equation}
where $f$ is the Fermi-Dirac distribution function of energy $\omega$ and $\rho$ is the density of states in the leads.
Applying the optical theorem and using the transition amplitude~\cite{Glazman} $\mathcal{A}_{|k'\lambda',\sigma'\rangle\leftarrow|k\lambda,\sigma\rangle}$ of reservoir states due to the scattering by the TQD, we write
\begin{eqnarray}
\textrm{Im} \mathcal{T}_\lambda \propto \sum_{\sigma,\sigma',\lambda'}|\mathcal{A}_{|k'\lambda',\sigma'\rangle\leftarrow|k\lambda,\sigma\rangle}|^2.\nonumber
\end{eqnarray}
After some straightforward calculations and by putting the coefficients of Eq. (1) and Eq. (4), we find
\begin{eqnarray}
\textrm{Im} \mathcal{T}_\lambda \propto \frac{1}{2}J_{z\lambda}^2(\omega) + J_+^2(\omega) \propto (J_z(\omega) + m_\lambda)^2 + n_\lambda \nonumber
\end{eqnarray}
where $m_\lambda=\chi^2\zeta_\lambda/(2l^2+\chi^2)$, $n_\lambda=(2l^2\zeta_\lambda^2\chi^2-4l^4\Delta_J^2-2l^2\Delta_J^2\chi^2)/(2l^2+\chi^2)^2$, $\zeta_{A,B}=l(J_{zA,zB}+J_{zC})/3+(l-1)/\rho$, $\zeta_C=-(J_{zA}+J_{zC})/l+(l-1)/\rho$, and $\chi=\sqrt{L/(2\pi a)}$. Note that $m_\lambda$ and $n_\lambda$ are independent on $T$, because $J_{zA}+J_{zC}$ is a coefficient of the marginal term in Poorman's scaling. Using an expression, $J_z(T)=\Delta_J(1+(T/T_{K})^{-4\rho \Delta_J})/(1-(T/T_{K})^{-4\rho \Delta_J})$ from Poorman's scaling, we obtain 
Eq. (5).

Finally, we discuss the derivation of Eq. (6).
The strong-coupling fixed point Hamiltonian for the usual spin-1/2 Kondo model is written as~\cite{Nozieres} 
\begin{equation}
\makeatletter 
\renewcommand{\theequation}{S\@arabic\c@equation}
\makeatother
\mathcal{H} = \mathcal{H}_\textrm{L} - \frac{1}{2\pi\rho T_K} \sum_{\sigma=\uparrow,\downarrow;k,k'}(\epsilon_k + \epsilon_{k'})  f^\dagger_{\sigma k} f_{\sigma k'} + \frac{1}{\pi\rho^2 T_K}\sum_{k_1,k_2,k_3,k_4} f_{\uparrow k_1}^\dagger f_{\uparrow k_2} f_{\downarrow k_3}^\dagger f_{\downarrow k_4}, \label{aaaaa}
\end{equation}
where $f_{k \sigma}^\dagger$ creates an electron with momentum $k$ and spin $\sigma$ in a reservoir.
In our case, $f_{k \sigma}$ is the pseudofermion operator defined in Eq. (4).
Using Eq.~(S\ref{F-operator}), we rewrite the Hamiltonian in Eq.~(S\ref{aaaaa})  in terms of $\phi_\sigma$ and $N_\sigma$.
Then, using the transformation in Eqs. (2) and (3), the Hamiltonian is expressed in terms of the operators of lead $\lambda$, which leads to Eq. (6).

{}